# High-resolution radio observations of massive protostars in the SARAO MeerKAT Galactic Plane Survey

W.O Obonyo,[1]⋆ M.G Hoare[2], S.L Lumsden[2], J. O. Chibueze[1], C.J. Ugwu[1]

[1] *Department of Mathematical Sciences, University of South Africa, Cnr Christian de Wet Rd and Pioneer Avenue, Florida Park, 1709, Roodepoort, South Africa.*
[2] *School of Physics and Astronomy, The University of Leeds, Woodhouse Lane, Leeds LS2 9JT, United Kingdom.*

**ABSTRACT**

We present high-resolution observations made with the Australia Telescope Compact Array (ATCA) at 5.5 GHz and 9.0 GHz of a sample of twenty-eight massive protostars. These protostars were initially detected at radio wavelengths in the MeerKAT Galactic Plane Survey at 1.3 GHz, where they were unresolved with an angular resolution of approximately 8 arcseconds. The resolution of the ATCA observations at 5.5 GHz and 9.0 GHz are significantly higher at, 2.0″ and 1.2″, respectively. The highest angular resolution corresponds to a linear resolution of 1920 AU for our nearest object. This improvement in resolution enabled us to resolve the components of extended emission and to more accurately determine the nature of the objects. Approximately 80% of the detections were classified as jets with 68% of the jets found to be associated with non-thermal emission.

## 1 Introduction

Massive protostars form through the accretion of materials that are often accompanied by the ejection of a fraction of the accreted material, closely linking the jet-driving phase to star formation processes (Carrasco-González et al. 2021). This ejection has been proposed to be either magnetically (Pudritz et al. 2007; Shu et al. 1994b; Carrasco-González et al. 2010) or radiatively (Kuiper et al. 2015; Carrasco-González et al. 2015) launched, accelerated and collimated. The ejections play a crucial role as they drive mass away from the accretion disks, thereby conserving angular momentum and facilitating further accretion onto the protostar (Lovelace et al. 1995). Despite their astrophysical significance, the fundamental processes underlying the ejection of mass to form these jets remain unresolved. This is mainly due to the limitations in current observations, which lack the angular resolution required to resolve the inner 100AU of the jets, where launching and collimation occur (Carrasco-González et al. 2021). Indeed, the most current observational representation of a massive protostellar jet is that of the Cep A HW2 radio jet (Carrasco-González et al. 2021), where the outflowing material is composed of two components; a wide-angle spherical wind launched from the protostellar disk, and a highly collimated jet emanating from approximately 20–30 AU from the protostar.

The protostars form in dusty environments where they are typically deeply embedded in their natal clouds. Their jets serve as the earliest indicators of massive star formation at radio wavelengths. These radio jets are now known to have components that may exhibit positive or negative spectral indices, depending on the position of the radio source relative to the central driving source (Garay et al. 2003; Guzmán et al. 2012; Moscadelli et al. 2016; Anglada et al. 2018; Purser et al. 2016; Obonyo et al. 2019). The spectral index, $\alpha$, is defined by the relation $S_\nu \propto \nu^\alpha$, where $S_\nu$ is the flux density at frequency $\nu$. Generally, the region associated with a luminous infrared source exhibit a positive index, while the more separated lobes may display either positive or negative spectral indices (Purser et al. 2016, Obonyo et al. 2019, Sanna et al. 2019, Obonyo et al. 2021) at centimeter wavelengths. The thermal component of the jet that is associated with an infrared source is known as a core, while the separated parts are called lobes (Purser et al. 2016, Obonyo et al. 2019, Sanna et al. 2019).

Moreover, it is increasingly becoming evident that about 40-50% massive protostellar jets are associated with non-thermal emission, with the rest emitting free-free emission. This split could indicate an evolutionary sequence, consistent with the current theory where the envelope of an evolving massive protostar undergoes a convective phase before becoming fully radiative (Hosokawa & Omukai 2009). While theories of massive star formation typically suggest the formation of jets before HII regions (Tanaka et al. 2016), the observation by Guzmán et al. (2016) manifested a contemporaneous jet and HII region in the same object.

Advances in telescope sensitivity and resolution have significantly improved the detection of massive protostars, including the phases exhibiting weak radio emissions, such as the jet driving phase (Purser et al. 2016, Obonyo et al. 2019, Sanna et al. 2019). Nonetheless, previous Galactic Plane surveys e.g Wood & Churchwell (1989), Urquhart et al. (2009), Irabor et al. (2023) with typical sensitivities $\simeq 0.32$ mJy beam$^{-1}$, 0.2 mJy beam$^{-1}$ and 0.11 mJy beam$^{-1}$, respectively, were not sensitive enough to detect large numbers of the MYSO jets. Notably, the SARAO MeerKAT Galactic Plane Survey (SMGPS; Goedhart et al. 2024), conducted at 1.3 GHz with a sensitivity $\simeq 0.01-0.02$ mJy beam$^{-1}$ detected a population of radio weak massive young stellar objects (MYSOs) in the southern hemisphere

⋆ E-mail: obonyow@gmail.com

(Obonyo et al. 2024). The low resolution of the observation, however, did not resolve most of the jets, which were identified statistically (Lumsden et al. 2013) using their mid- and near-infrared colors and through visual inspection.

While the infrared studies of MYSOs have the potential to identify most of these objects, there is a possibility of contamination from HII regions. However, these objects can be resolved using the Australian Compact Array Telescope (ATCA) at higher frequencies, revealing details of their morphologies. For instance, Purser et al. (2016) reported resolved jets in their ATCA observations, where radio emission manifested the separation of some ionized jets' cores from their lobes. In this work, we report a simultaneous observation of the objects with ATCA at both 5.5 GHz and 9.0 GHz to reveal their morphologies and to constrain their spectral indices.

## 2 Observation and Data

### 2.1 Sample selection

Twenty-seven of the MYSO in Obonyo et al. (2024) had not been studied in previous higher-frequency radio studies, e.g, Helfand et al. 2006 and Urquhart et al. 2007). This was either due to their low declinations, $\delta \leq 40°$, or their low surface brightness at radio wavelength, and were selected for the study. Indeed many of the objects lie within the CORNISH-South survey area but were too weak to detect (Irabor et al. 2023). Additionally, G287.3716+00.6444 which was classified by Purser et al. (2016) as an HII region was included in the sample to search for variability. All the selected objects were unresolved in the MeerKAT observation and did not reveal any clear jet morphology even though their IR images display bipolar/monopolar nebulae, indicative of outflow activity. Table 1 shows a list of the selected MYSOs; columns 1 - 9 show the RMS† name, associated IRAS‡ object, right ascension, declination, distance, bolometric luminosities, rms noise at 5.5 GHz, rms noise at 9.0 GHz, and the integrated fluxes of the radio sources at 1.3 GHz (Obonyo et al. 2024), respectively.

### 2.2 ATCA observation

The observations were made between 21$^{st}$ November 2022 and 16$^{th}$ January 2023 using the ATCA's 6C array configuration at 5.5, and 9.0 GHz. The observations, done under the project code C3508, made use of all the six antennas of the configuration where the shortest and longest baselines are 153 and 6000 m, corresponding to angular resolutions of 2″ and 1″ at 5.5 GHz (C-band) and 9.0 GHz (C-band), respectively. The bandwidth for each of the frequencies of observation was 2 GHz. The total on-source integration time at each frequency was 72 minutes, resulting from 12 uv-cuts spread over a 12-hour track to provide good uv-coverage. The sources were observed in groups of four sources, for 6 minutes each, followed by a one-minute observation of a phase calibrator to correct for atmospheric and instrumental effects on the amplitudes and phases of the data. This was repeated for a further set of three sources.

Calibration and data reduction were performed using the MIRIAD§ astronomy software (Sault et al. 1995). Whenever a radio source with SNR ≥ 10 was present within the primary beam of the telescope, we used the source to perform self-calibration on the phases of the data. Maps of the fields were made by performing Fourier transform on the visibilities with robust weighting (Briggs 1995) of robustness parameter r = 0.5, obtaining synthesized beams whose typical FWHM¶ are 2.1″ by 1.8″ and 1.3″ by 1.1″ at 5.5, and 9.0 GHz, respectively. The dirty maps obtained were deconvolved using MIRIAD's MFCLEAN task. The average root-mean-square (rms) noise of the deconvolved maps typically lies within the range 22-50 and 15-50 μJybeam$^{-1}$ at 5.5 and 9.0 GHz, respectively. However, in a few of the fields, e.g. in G345.5043+00.3480's field, the rms is almost a factor of a hundred higher than the typical noise of the observation. This is largely due to the presence of bright extended emission that lowers the dynamic range in such a field. The flux density and bandpass calibrator for the observation was 1934-638 (3C84).

The astrometry of the MeerKAT and ATCA datasets were examined as both were used in the analysis. Goedhart et al. (2024) estimated the positional accuracy of MeerKAT observations by comparing the coordinates of point sources to their corresponding positions in the CORNISH and CORNISH-South surveys (Hoare et al. 2012; Irabor et al. 2023). They reported a systematic positional error, with angular offsets ranging from 0 to 2.1″, as illustrated in Figure 6 in Goedhart et al. (2024). This suggests the potential for positional discrepancies of up to 2.1″ between MeerKAT and ATCA data.

## 3 Radio properties of the massive protostars manifesting low surface-brightness in the SARAO MeerKAT GPS

We detected centimetre radio continuum emission from nineteen sources at 5.5 GHz and twenty sources at 9.0 GHz at 3σ level and above. The positions of each of the detected sources were derived from 2D Gaussian fits at each frequency and found to be consistent at all frequencies within 0.02−0.08 arcseconds, with the largest difference registered in sources where structures are resolved at one frequency and not the other. This range of uncertainties is also in agreement with the accuracies stated in the ATCA information manual. The positions of the sources at 5.5 GHz and 9.0 GHz are listed in Table 2 with the sources which were not detected at the frequencies of observation represented by dashes.

The other observed parameters of the sources; flux densities, deconvolved angular sizes, position angles (PA), and peak fluxes of the MYSOs at 5.5 and 9.0 GHz are presented in Table 3. The fluxes for each of the sources were estimated in two methods; using the MIRIAD's task, imfit, that fits a 2-dimensional Gaussian function to the detections and secondly by summing up the emission within a region encompassing the source. For sources with irregular or non-Gaussian morphologies, the fluxes were estimated by calculating the total flux density within a box enclosing the source. In most cases, the source morphologies were well estimated by Gaussian fits, with the major and minor axes listed for each source being the deconvolved dimensions derived from the fits.

The range, average and standard deviation of the fluxes at 5.5 GHz are 0.07−3.86 mJy, 0.67 mJy and 0.84 mJy, respectively, with a majority of the sources having integrated flux densities $S_\nu < 1.5$ mJy. Two of the sources, G287.3716+00.6444-A, and G340.7455-01.0021, however, have flux densities, $S_\nu > 3$ mJy, i.e, 3.86±0.21 mJy and 3.13±0.26 mJy, respectively. A comparable trend is seen at 9.0 GHz, with the same objects manifesting the highest fluxes again at 9.0 GHz.

---

† http://rms.leeds.ac.uk/cgi-bin/public/RMS_DATABASE.cgi
‡ Infrared Astronomical Satellite
§ Multichannel Image Reconstruction, Image Analysis and Display

¶ Full Width at Half Maximum

Table 1. Table showing object names, associated IRAS objects, positions (Goedhart et al. 2024) where observation pointing centres were set to, distance from the Sun, and Bolometric luminosities which were taken from Lumsden et al. (2013) catalogue. Local rms noise levels of the maps and the integrated fluxes of the sources in the MeerKAT's 1.3 GHz observation are also shown.

| RMS Name | Associated IRAS Object | RA (J2000) | Dec (J2000) | D (kpc) | $L_{Bol}$ ($L_\odot$) | rms (5.5 GHz) ($\mu Jy/beam$) | rms (9.0 GHz) ($\mu Jy/beam$) | Int. Flux (mJy) (1.3 GHz) |
|---|---|---|---|---|---|---|---|---|
| G270.8247-01.1112 | IRAS 09088-4929 | $09^h10'30.89''$ | $-49°41'29.8''$ | 6.3 | 7100 | 15 | 21 | 0.09±0.03 |
| G271.2225-01.7712 | IRAS 09075-5013 | $09^h09'11.16''$ | $-50°25'55.3''$ | 8.6 | 4800 | 25 | 24 | 0.80±0.08 |
| G281.2206-01.2556 | IRAS 09598-5632 | $10^h01'33.78''$ | $-56°47'18.2''$ | 7.0 | 3900 | 25 | 24 | 0.08±0.02 |
| G282.8969-01.2727 | – | $10^h11'31.60''$ | $-57°47'03.7''$ | 7.0 | 17000 | 38 | 27 | 0.75±0.05 |
| G287.3716+00.6444 | IRAS 10460-5811 | $10^h48'04.55''$ | $-58°27'01.5''$ | 4.5 | 18000 | 27 | 37 | 4.22±0.38 |
| G287.8768-01.3618 | IRAS 10423-6011 | $10^h44'17.93''$ | $-60°27'46.0''$ | 6.1 | 19000 | 28 | 26 | 17.4±0.3 |
| G289.0543+00.0223 | – | $10^h57'42.91''$ | $-59°44'58.4''$ | 7.3 | 2900 | 27 | 24 | 0.30±0.04 |
| G293.0352-00.7871 | IRAS 11233-6143 | $11^h25'33.22''$ | $-61°59'51.3''$ | 3.3 | 1900 | 23 | 17 | 0.78±0.05 |
| G297.1390-01.3510 | IRAS 11563-6320 | $11^h58'54.99''$ | $-63°37'47.8''$ | 8.9 | 12000 | 24 | 17 | 1.34±0.16 |
| G300.7221+01.2007 | IRAS 12300-6119 | $12^h32'50.15''$ | $-61°35'27.0''$ | 4.3 | 1600 | 33 | 17 | 0.38±0.02 |
| G301.0130+01.1153 | – | $12^h35'14.47''$ | $-61°41'46.8''$ | 4.3 | 2400 | 85 | 48 | 0.21±0.03 |
| G304.7592-00.6299 | IRAS 13046-6310 | $13^h07'47.42''$ | $-63°26'37.0''$ | 11.2 | 4100 | 32 | 19 | 0.63±0.02 |
| G309.9796+00.5496 | IRAS 13475-6115 | $13^h51'02.72''$ | $-61°30'14.1''$ | 3.5 | 7600 | 64 | 21 | 0.80±0.03 |
| G314.3197+00.1125 | IRAS 14227-6025 | $14^h26'26.28''$ | $-60°38'31.5''$ | 3.6 | 38000 | 26 | 19 | 0.41±0.03 |
| G328.2523-00.5320A | IRAS 15541-5349 | $15^h57'59.82''$ | $-53°58'00.4''$ | 2.9 | 8000 | 220 | 95 | 0.92±0.03 |
| G328.3442-00.4629 | IRAS 15543-5342 | $15^h58'09.62''$ | $-53°51'18.2''$ | 2.9 | 2400 | 305 | 57 | 0.93±0.03 |
| G328.9842-00.4361 | IRAS 15574-5316 | $16^h01'19.12''$ | $-53°25'04.2''$ | 4.7 | 1800 | 23 | 17 | 0.63±0.03 |
| G329.0663-00.3081 | IRAS 15573-5307 | $16^h01'09.93''$ | $-53°16'02.3''$ | 11.6 | 36000 | 24 | 16 | 0.33±0.02 |
| G334.7302+00.0052 | IRAS 16223-4901 | $16^h26'04.65''$ | $-49°08'41.8''$ | 2.5 | 3800 | 27 | 27 | 0.12±0.01 |
| G335.0611-00.4261A | IRAS 16256-4905 | $16^h29'22.98''$ | $-49°12'27.0''$ | 2.8 | 4200 | 37 | 20 | 1.18±0.16 |
| G338.2801+00.5419A | IRAS 16344-4605 | $16^h38'09.11''$ | $-46°11'04.0''$ | 4.1 | 5100 | 26 | 16 | 0.43±0.08 |
| G338.4712+00.2871 | IRAS 16363-4606 | $16^h39'58.90''$ | $-46°12'36.4''$ | 13.1 | 170000 | 60 | 84 | 0.80±0.03 |
| G339.3316+00.0964 | IRAS 16404-4535 | $16^h44'04.39''$ | $-45°41'27.1''$ | 13.1 | 17000 | 52 | 27 | 2.12±0.04 |
| G340.7455-01.0021 | – | $16^h54'04.05''$ | $-45°18'50.0''$ | 2.6 | 5300 | 978 | 877 | 1.63±0.05 |
| G345.2619-00.4188A | IRAS 17033-4119 | $17^h06'50.54''$ | $-41°23'46.4''$ | 2.7 | 1200 | 23 | 30 | 0.27±0.03 |
| G345.5043+00.3480 | IRAS 17008-4040 | $17^h04'22.87''$ | $-40°44'23.5''$ | 2.0 | 100000 | 1923 | 1937 | 1.77±0.07 |
| G345.7172+00.8166A | IRAS 16596-4012 | $17^h03'06.40''$ | $-40°17'08.7''$ | 1.6 | 1100 | 27 | 21 | 0.68±0.03 |
| G345.9561+00.6123 | IRAS 17012-4009 | $17^h04'43.00''$ | $-40°13'13.4''$ | 2.5 | 2400 | 23 | 19 | < 0.09 |

**Table 2.** Coordinates of sources detected at 5.5 GHz, 9.0 GHz, or both frequencies. A − indicates non-detection at a given frequency or cases where the source's position could not be determined due to being resolved. The table also includes the spectral indices of the detected sources.

| RMS Name | Position at C-band | | Position at X-band | |
|---|---|---|---|---|
| | RA(J2000) | Dec(J2000) | RA(J2000) | Dec(J2000) |
| G270.8247-01.1112 | $09^h 10' 30.77''$ | -049°41′27.06″ | − | − |
| G271.2225-01.7712 | $09^h 09' 11.17''$ | -050°25′55.04″ | $09^h 09' 11.17''$ | -050°25′55.08″ |
| G281.2206-01.2556 | − | − | − | − |
| G282.8969-01.2727-A | $10^h 11' 31.56''$ | -057°47′03.95″ | − | − |
| G282.8969-01.2727-B | $10^h 11' 31.71''$ | -057°47′04.99″ | − | − |
| G287.3716+00.6444-A | $10^h 48' 04.66''$ | -058°27′02.10″ | $10^h 48' 04.65''$ | -058°27′02.21″ |
| G287.3716+00.6444-B | $10^h 48' 05.19''$ | -058°26′57.17″ | $10^h 48' 05.32''$ | -058°26′57.78″ |
| G287.8768-01.3618 | $10^h 44' 17.91''$ | -060°27′45.08″ | $10^h 44' 17.94''$ | -060°27′45.33″ |
| G289.0543+00.0223 | − | − | $10^h 57' 42.89''$ | -059°44′57.65″ |
| G293.0352-00.7871 | − | − | $11^h 25' 33.14''$ | -061°59′51.55″ |
| G297.1390-01.3510 | $11^h 58' 54.95''$ | -063°37′47.67″ | − | − |
| G297.1390-01.3510A | − | − | $11^h 58' 54.86''$ | -063°37′47.90″ |
| G297.1390-01.3510B | − | − | $11^h 58' 54.90''$ | -063°37′46.41″ |
| G300.7221+01.2007 | − | − | − | − |
| G301.0130+01.1153 | − | − | − | − |
| G304.7592-00.6299 | $13^h 07' 47.42''$ | -063°26′36.76″ | $13^h 07' 47.40''$ | -063°26′37.08″ |
| G309.9796+00.5496 | − | − | − | − |
| G309.9796+00.5496A | $13^h 51' 02.70''$ | -061°30′13.43″ | $13^h 51' 02.66''$ | -061°30′13.63″ |
| G309.9796+00.5496B | $13^h 51' 03.00''$ | -061°30′13.15″ | $13^h 51' 03.06''$ | -061°30′12.71″ |
| G314.3197+00.1125-A | $14^h 26' 26.26''$ | -060°38′31.29″ | $14^h 26' 26.30''$ | -060°38′31.31″ |
| G314.3197+00.1125-B | $14^h 26' 26.12''$ | -060°38′29.44″ | − | − |
| G314.3197+00.1125-C | $14^h 26' 26.66''$ | -060°38′39.88″ | $14^h 26' 26.64''$ | -060°38′39.91″ |
| G328.2523-00.5320A | − | − | − | − |
| G328.2523-00.5320A-A | $15^h 57' 59.46''$ | -053°58′01.56″ | $15^h 57' 59.44''$ | -053°58′01.70″ |
| G328.2523-00.5320A-B | $15^h 57' 59.82''$ | -053°57′59.99″ | $15^h 57' 59.83''$ | -053°58′00.23″ |
| G328.2523-00.5320A-C | $15^h 57' 59.05''$ | -053°58′03.39″ | $15^h 57' 59.05''$ | -053°58′03.12″ |
| G328.3442-00.4629 | − | − | − | − |
| G328.9842-00.4361 | $16^h 01' 19.09''$ | -053°25′03.76″ | $16^h 01' 19.12''$ | -053°25′04.06″ |
| G329.0663-00.3081 | − | − | − | − |
| G334.7302+00.0052 | $16^h 26' 04.65''$ | -049°08′42.03″ | $16^h 26' 04.66''$ | -049°08′42.22″ |
| G335.0611-00.4261A-(A+B) | $16^h 29' 23.16''$ | -049°12′28.25″ | − | − |
| G335.0611-00.4261A-A | − | − | $16^h 29' 23.19''$ | -049°12′28.84″ |
| G335.0611-00.4261A-B | − | − | $16^h 29' 23.11''$ | -049°12′27.30″ |
| G335.0611-00.4261A-C | $16^h 29' 23.13''$ | -049°12′31.84″ | $16^h 29' 23.09''$ | -049°12′31.40″ |
| G338.2801+00.5419A-A | $16^h 38' 08.97''$ | -046°10′59.00″ | $16^h 38' 08.96''$ | -046°10′58.81″ |
| G338.2801+00.5419A-B | $16^h 38' 09.06''$ | -046°11′03.43″ | $16^h 38' 09.07''$ | -046°11′03.27″ |
| G338.4712+00.2871 | $16^h 39' 58.59''$ | -046°12′37.90″ | $16^h 39' 58.59''$ | -046°12′37.96″ |
| G339.3316+00.0964 | $16^h 44' 04.36''$ | -045°41′27.51″ | $16^h 44' 04.36''$ | -045°41′27.59″ |
| G340.7455-01.0021 | $16^h 54' 04.03''$ | -045°18′49.64″ | $16^h 54' 04.05''$ | -045°18′50.02″ |
| G345.2619-00.4188A | $17^h 06' 50.76''$ | -041°23′44.39″ | − | − |
| G345.5043+00.3480 | − | − | − | − |
| G345.7172+00.8166A | $17^h 03' 06.37''$ | -040°17′08.81″ | $17^h 03' 06.39''$ | -040°17′08.81″ |
| G345.9561+00.6123 | − | − | − | − |

Table 3. A table of massive protostars and their components detected in the ATCA observations. Their deconvolved sizes, integrated fluxes, and peak fluxes at both 5.5 GHz and 9.0 GHz are shown. — indicates a non-detection, combined emission, or when a source is split into components at a given frequency, making it impossible to determine its parameters at that frequency.

| RMS Name | C-band Parameters | | | | | | X-band Parameters | | | | | | $\alpha$ | Object |
|---|---|---|---|---|---|---|---|---|---|---|---|---|---|---|
| | C-band Size | | PA | Int. Flux | Peak Flux | | X-band Size | | PA | Int. Flux | Peak Flux | | | |
| | $\theta_{maj}$ (″) | $\theta_{min}$ (″) | (°) | (mJy) | (mJy/beam) | | $\theta_{maj}$ (″) | $\theta_{min}$ (″) | (°) | (mJy) | (mJy/beam) | | | |
| G270.8247−01.1112 | 2.38±0.92 | 0.65±0.91 | 54±43 | 0.13±0.04 | 0.08±0.02 | | — | — | — | — | <0.06 | | <−1.57 | Lobe |
| G271.2225−01.7712 | 2.86±0.32 | 1.64±0.11 | 177±5 | 0.35±0.05 | 0.24±0.02 | | <1.19 | <1.11 | <169 | 0.28±0.04 | 0.25±0.02 | | −0.45±0.09 | Jet |
| G281.2206−01.2556 | — | — | — | — | <0.08 | | — | — | — | — | <0.08 | | — | Core |
| G282.8969−01.2727-A | — | — | — | 0.09±0.03 | 0.07±0.01 | | — | — | — | — | <0.08 | | <−0.24 | Core |
| G282.8969−01.2727-B | — | — | — | 0.07±0.02 | 0.06±0.01 | | — | — | — | — | <0.08 | | <0.27 | Lobe |
| G287.3716+00.6444-A | 1.83±0.20 | 1.31±0.22 | 137±19 | 3.86±0.21 | 2.28±0.09 | | 1.42±0.13 | 1.30±0.15 | 112±79 | 4.43±0.28 | 1.90±0.09 | | 0.28±0.04 | Core |
| G287.3716+00.6444-B | 3.94±0.64 | 2.04±0.47 | 66±11 | 0.67±0.10 | 0.20±0.02 | | 1.99±0.22 | 1.25±0.17 | 165±11 | 0.48±0.05 | 0.14±0.02 | | −0.68±0.08 | Lobe |
| G287.8768−01.3618 | 1.82±0.32 | 0.19±0.53 | 148±10 | 0.58±0.06 | 0.42±0.03 | | 2.28±0.25 | 0.45±0.24 | 161±3 | 0.93±0.09 | 0.40±0.03 | | 0.96±0.06 | Jet |
| G289.0543+00.0223 | — | — | — | — | <0.09 | | 1.32±0.41 | 1.08±0.38 | 83±91 | 0.22±0.04 | 0.11±0.02 | | >1.81 | HII-region |
| G293.0352−00.7871 | — | — | — | — | <0.08 | | <0.81 | <0.3 | <166 | 0.14±0.02 | 0.12±0.02 | | >1.14 | Unknown |
| G297.1390−01.3510 | 4.76±0.54 | 1.06±0.37 | 77±3 | 0.63±0.07 | 0.19±0.02 | | — | — | — | — | — | | −1.25±0.08 | Jet |
| G297.1390−01.3510A | — | — | — | — | — | | 1.65±0.23 | 1.21±0.34 | 109±35 | 0.23±0.04 | 0.09±0.01 | | — | Lobe |
| G297.1390−01.3510B | — | — | — | — | — | | 1.47±0.31 | 0.13±0.46 | 78±10 | 0.12±0.02 | 0.07±0.01 | | — | Unknown |
| G300.7221+01.2007 | — | — | — | — | <0.09 | | — | — | — | — | <0.05 | | — | — |
| G301.0130+01.1153 | — | — | — | — | <0.26 | | — | — | — | — | <0.14 | | — | — |
| G304.7592−00.6299 | 2.51±0.38 | 1.00±0.48 | 98±9 | 0.61±0.07 | 0.31±0.03 | | 1.53±0.33 | 0.82±0.31 | 140±21 | 0.38±0.06 | 0.19±0.02 | | −0.93±0.09 | HII-region |
| G309.9796+00.5496 | — | — | — | 0.46±0.10 | — | | — | — | — | 0.59±0.09 | — | | 0.51±0.12 | Jet |
| G309.9796+00.5496A | — | — | — | 0.20±0.08 | 0.16±0.04 | | 2.70±0.67 | 1.00±0.47 | 100±10 | 0.30±0.07 | 0.09±0.02 | | 0.82±0.20 | Core |
| G309.9796+00.5496B | 1.74±0.59 | 0.55±0.64 | 30±47 | 0.26±0.06 | 0.19±0.03 | | 1.83±0.42 | 0.94±0.39 | 53±18 | 0.29±0.05 | 0.11±0.02 | | 0.22±0.13 | Lobe |
| G314.3197+00.1125-A | — | — | — | 0.08±0.02 | 0.06±0.01 | | 1.28±0.13 | 0.26±0.19 | 116±6 | 0.17±0.02 | 0.11±0.01 | | 1.53±0.12 | Jet |
| G314.3197+00.1125-B | — | — | — | 0.07±0.02 | 0.04±0.01 | | — | — | — | — | <0.05 | | <−0.68 | Lobe |
| G314.3197+00.1125-C | — | — | — | 0.12±0.02 | 0.05±0.01 | | — | — | — | 0.09±0.03 | 0.08±0.02 | | −0.58±0.16 | Unknown |
| G328.2523−00.5320A-A | — | — | — | 0.48±0.08 | 0.45±0.04 | | — | — | — | 0.42±0.07 | 0.34±0.05 | | −0.27±0.10 | Lobe |
| G328.2523−00.5320A-B | — | — | — | 0.33±0.04 | 0.29±0.03 | | — | — | — | 0.54±0.09 | 0.46±0.06 | | 1.00±0.09 | Core |
| G328.2523−00.5320A-C | — | — | — | 0.23±0.07 | 0.21±0.02 | | 1.53±0.32 | 0.79±0.33 | 34±12 | 0.61±0.08 | 0.31±0.03 | | 1.98±0.14 | Unknown |
| G328.3442−00.4629 | — | — | — | — | <0.63 | | — | — | — | <0.17 | — | | — | — |
| G328.9842+00.4361 | 1.93±0.57 | 0.24±0.69 | 143±21 | 0.22±0.04 | 0.16±0.02 | | 2.24±0.70 | 0.32±0.49 | 141±13 | 0.24±0.06 | 0.12±0.02 | | 0.18±0.13 | Jet |
| G329.0663−00.3081 | — | — | — | — | <0.07 | | — | — | — | — | <0.05 | | — | — |
| G334.7302+00.0052 | 2.17±0.15 | 1.69±0.09 | 1±9 | 0.31±0.04 | 0.30±0.02 | | 1.12±0.28 | 0.75±0.33 | 150±51 | 0.55±0.05 | 0.38±0.02 | | 1.20±0.07 | Jet |
| G335.0611−00.4261A-(A+B) | 3.13±0.44 | 2.15±0.35 | 160±23 | 0.76±0.08 | 0.30±0.02 | | — | — | — | — | — | | −0.38±0.06 | Jet |
| G335.0611−00.4261A-A | — | — | — | — | — | | 0.73±0.26 | 0.71±0.38 | 32±83 | 0.25±0.03 | 0.18±0.02 | | — | Lobe |
| G335.0611−00.4261A-A | — | — | — | — | — | | 2.04±0.24 | 0.86±0.26 | 66±8 | 0.38±0.04 | 0.15±0.02 | | — | Core |
| G335.0611−00.4261A-C | — | — | — | 0.11±0.02 | 0.09±0.01 | | — | — | — | 0.24±0.04 | 0.14±0.02 | | 1.58 | Unknown |
| G338.2801+00.5419A-(A+B) | — | — | — | 0.99±0.08 | — | | — | — | — | 0.74±0.04 | — | | — | Jet |
| G338.2801+00.5419A-A | 1.87±0.70 | 0.97±0.50 | 144±45 | 0.41±0.06 | 0.31±0.03 | | 1.18±0.28 | 0.43±0.20 | 167±17 | 0.28±0.03 | 0.20±0.01 | | −0.77±0.08 | Lobe |
| G338.2801+00.5419A-B | 2.12±0.47 | 1.46±0.43 | 165±87 | 0.58±0.06 | 0.35±0.03 | | 0.71±0.16 | 0.46±0.16 | 177±55 | 0.46±0.03 | 0.38±0.01 | | −0.46±0.05 | Core |
| G338.4712+00.2871 | 3.82±0.43 | 1.90±0.12 | 162±4 | 0.80±0.11 | 0.52±0.04 | | 2.32±0.43 | 1.12±0.12 | 152±5 | 0.45±0.11 | 0.32±0.05 | | −1.17±0.12 | Jet |
| G339.3316+00.0964 | 2.54±0.11 | 1.71±0.05 | 16±6 | 1.31±0.08 | 1.28±0.04 | | 0.97±0.19 | 0.69±0.20 | 21±46 | 1.60±0.11 | 1.14±0.05 | | 0.41±0.04 | Lobe |
| G340.7455−01.0021 | 1.95±0.18 | 1.15±0.06 | 171±4 | 3.13±0.26 | 3.54±0.23 | | 2.04±0.25 | 0.89±0.05 | 157±3 | 4.06±0.58 | 2.91±0.25 | | 0.53±0.08 | Core |
| G345.2619−00.4188A-A | 5.87±1.16 | 1.83±0.90 | 59±8 | 0.19±0.06 | 0.09±0.02 | | — | — | — | — | <0.09 | | <−1.52 | Lobe |
| G345.2619−00.4188A-B | — | — | — | 0.09±0.03 | 0.08±0.02 | | — | — | — | — | <0.09 | | <0.00 | Lobe |
| G345.5043+00.3480 | — | — | — | — | <16.80 | | — | — | — | — | <4.20 | | — | — |
| G345.7172+00.8166A | 2.56±0.21 | 2.32±0.18 | 139±32 | 0.26±0.04 | 0.20±0.02 | | 4.20±1.22 | 1.01±0.11 | 158±3 | 0.19±0.05 | 0.13±0.02 | | −0.64±0.13 | Jet |
| G345.9561+00.6123 | — | — | — | — | <7.10 | | — | — | — | — | <5.82 | | — | — |

## 3.1 Morphologies of the sources

The morphologies of radio jets from massive protostars can provide insights into the formation and evolution of the protostars and the associated jets (Cesaroni et al. 2018). The continuum images of the ATCA observations at 5.5 GHz and 9.0 GHz manifest a higher degree of detail of the unresolved 1.28 GHz MeerKAT images, revealing their morphologies in the majority of cases. Indeed, this is the first high-resolution radio observation of 27 massive protostars in the sample, contributing to the census of massive protostellar jets.

The twenty objects detected at either 5.5, 9.0 GHz, or both, as shown in Figure 1, exhibit a range of morphologies. Seven of these objects display multiple components, characteristic of a jet core-lobe(s) system. Examples include G282.8969-01.2727, G287.3716+00.6444, G309.9796+00.5496, G314.3197+00.1125, G328.2523-00.5320, G335.0611-00.4261A, and G338.2801+00.5419A. Another nine objects exhibit single extended elliptical emission, typical of jet-driven structures, such as G271.2225-01.7712, G287.8768-01.3618, G297.1390-01.3510, G328.9842-00.4361, G334.7302+00.0052, G338.4712+00.2871, G339.3316+00.0964, G340.7455-01.0021, and G345.7172+00.8166A. In contrast, objects such as G293.0352-00.7871 and G304.7592-00.6299 exhibit circular morphologies. The observed morphologies were key in identifying and classifying the objects, for example, objects with elliptical morphologies and multiple components are likely to host jet-driving sources.

Typically, collimated jets are manifested by narrow streams of radio emission that extend over large distances with minimal widening, often exhibiting internal shocks and knots along their lengths, e.g HH 80-81 (Rodríguez-Kamenetzky et al. 2017, Carrasco-González et al. 2010). While the objects in our sample do not manifest clear collimation over large distances, some of the jets show components that resemble core-lobe systems of jets, with others manifesting non-thermal emission suggesting that they host non-thermal lobes. Moreover, some of the sources show elliptical shapes of a few arcseconds, similar to the structures of cores of radio jets (Anglada et al. 2018).

## 3.2 Spectral Indices of the Objects

The spectral indices of the sources were calculated using their flux densities at 5.5 GHz and 9.0 GHz. To ensure consistency, the data at both frequencies were re-imaged by restricting the visibilities to baselines sensitive to emission from comparable angular scales. This was achieved by limiting the uv-range to a common range of 3– 90 k$\lambda$, applied to both frequency bands before estimating the indices. The full uv-coverage for the observations at 5.5 GHz and 9.0 GHz were 1.8−89.5 k$\lambda$ and 3.1−159.5 k$\lambda$, respectively. Notably, most fluxes from the common uv-range agree with those from the full uv-range within uncertainties. Flux densities at 1.3 GHz were excluded from the calculation of the indices due to their dissimilar uv-range compared to the other observations, and even if re-imaged, would not result in significant detections. However, for sources detected at all three frequencies, trends of their spectral energy distributions (SEDs) were inspected to help characterise the nature of their radio sources.

For radio sources detected at only one frequency, their spectral indices were constrained by setting upper or lower limits. These limits were determined using the measured flux at the detected frequency and three times the noise level ($3\sigma$) at the undetected frequency. Additionally, if a source was resolved into multiple components at a particular frequency, its spectral index was estimated by summing the fluxes of all components at that frequency.

Figures 2 presents a histogram illustrating the distribution of spectral indices for the detected sources and their resolved components, while Figure 3 displays the spectral energy distribution (SEDs) of the sources. Additionally, Table 3 lists the spectral indices, which range from approximately -1.3 to 1.7. This distribution indicates the presence of both thermal ($\alpha \geq -0.1$) and non-thermal ($\alpha < -0.1$) objects within the sample. The histogram shows two prominent peaks at $\alpha \simeq -0.57$ and 0.34, suggesting a mix of thermal and non-thermal radio emission in the jets. Notably, two sources exhibit highly negative spectral indices, e.g, $\alpha < -1.0$, atypical of synchrotron emission. Despite this, these sources, G297.1390-01.3510 and G338.4712+00.2871, were still classified as non-thermal jets or lobes.

Overall, the classification of the objects and their components based on their morphology, nature of emission, and association with IR and molecular emission are shown in Table 3. Out of the twenty-eight objects observed, twenty-two were detected. Of these, thirteen exhibited either non-thermal emission or were associated with non-thermal lobes, while some potential lobes were identified as thermal. Five radio sources were classified as unknowns, two as HII-regions, and the remaining were categorized as jets. These results suggest that non-thermal emission, and by extension magnetic fields, play a significant role in the formation of massive stars.

## 3.3 Nature of the objects

### 3.3.1 G270.8247-01.1112

The radio source associated with G270.8247-01.1112 is detected at both 1.28 GHz and 5.5 GHz but not 9.0 GHz, where the rms noise level is 0.021 mJy/beam. Although its spectral index is highly negative, <-1.57, its flux density at 1.28 GHz is lower than expected based on the SED generated from 5.5 and 9.0 GHz. This discrepancy suggests that the source may be a non-thermal lobe with a typical index of -0.6. Additionally, its position is offset to the north-west of IRS A, the potential jet driver, further supporting the interpretation that it is a potential lobe of a jet driven by IRS A.

G270.8247-01.1112 exhibits strong NIR CO emission (Ilee et al. 2013). Ilee et al. (2013) modelled the CO band-head emission using a disc model, establishing that it hosts a disc of size 0.1−1.6 au, inclined at 89°, i.e edge-on. Using data from the ALMA Evolutionary Study of High Mass Protocluster Formation in the Galaxy (ALMAGAL) survey (Wells et al. 2024), we extracted the $^{13}$CO (2 − 1) transition emission line at mm wavelength − 220.39868 GHz, revealing blue and red wings aligned in a SE-NW direction (see the red and blue contours on Figure 4).

Additionally, the millimeter continuum emission shows a single core coincident with IRS A. The similarity between the alignment of the radio source with IR source A and the direction of the molecular outflow suggests that the radio emission originates from a lobe of an ionized jet aligned with this outflow and driven by IRS A.

### 3.3.2 G271.2225-01.7712

G271.2225-01.7712 has a spectral index of -0.45±0.09, indicating non-thermal emission characteristic of radiation from non-thermal lobes of protostellar jets. The morphology of the object at both 5.5 GHz and 9.0 GHz, generated from data of similar uv-range, depicts a jet-like structure that is aligned in a N-S direction. However, no clear elongation is observed in the 2MASS and VVV (Minniti et al. 2010) infrared emissions. Similarly, Navarete et al. (2015) did not detect both molecular hydrogen emission at 2.12 μm, which is typically associated with shocks, or extended continuum infrared emission.

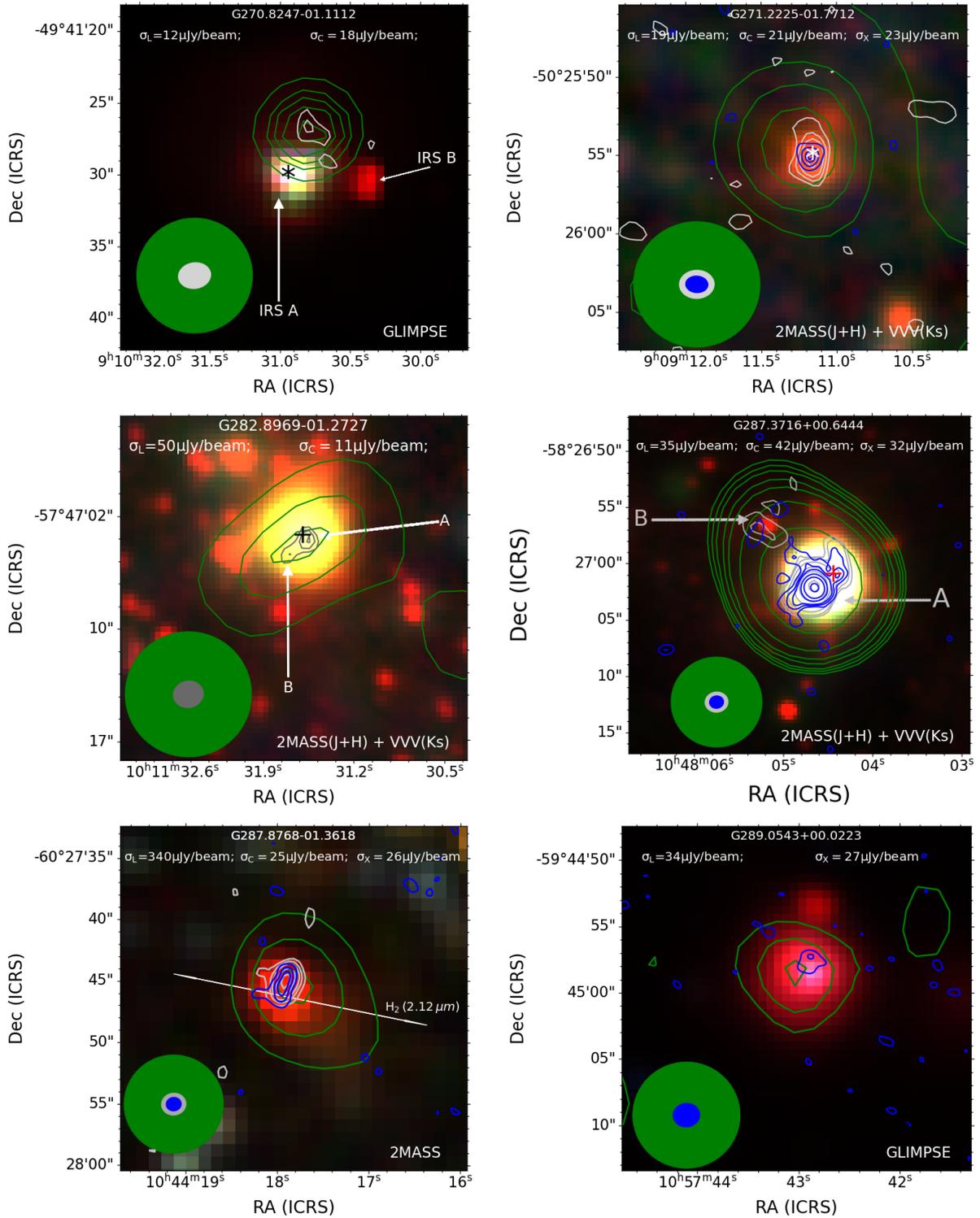

**Figure 1.** IR images of the sources together with their L-, C- and X-band contours of levels 3σ, 5σ, 7σ, 11σ, 15σ, 25σ and 50σ,... shown in green, gray and blue colours respectively. The synthesized beams for L-, C-, and X-bands are shown in the lower left corner of each frame. The locations of mm-cores, and methanol masers within the vicinity of the massive protostars are marked with asterisks(*) and plus(+) signs respectively. The sources of the infrared images are indicated in the lower right corners of the plots. GLIMPSE images are made from composite images of 3.6, 4.5, and 8.0 μm.

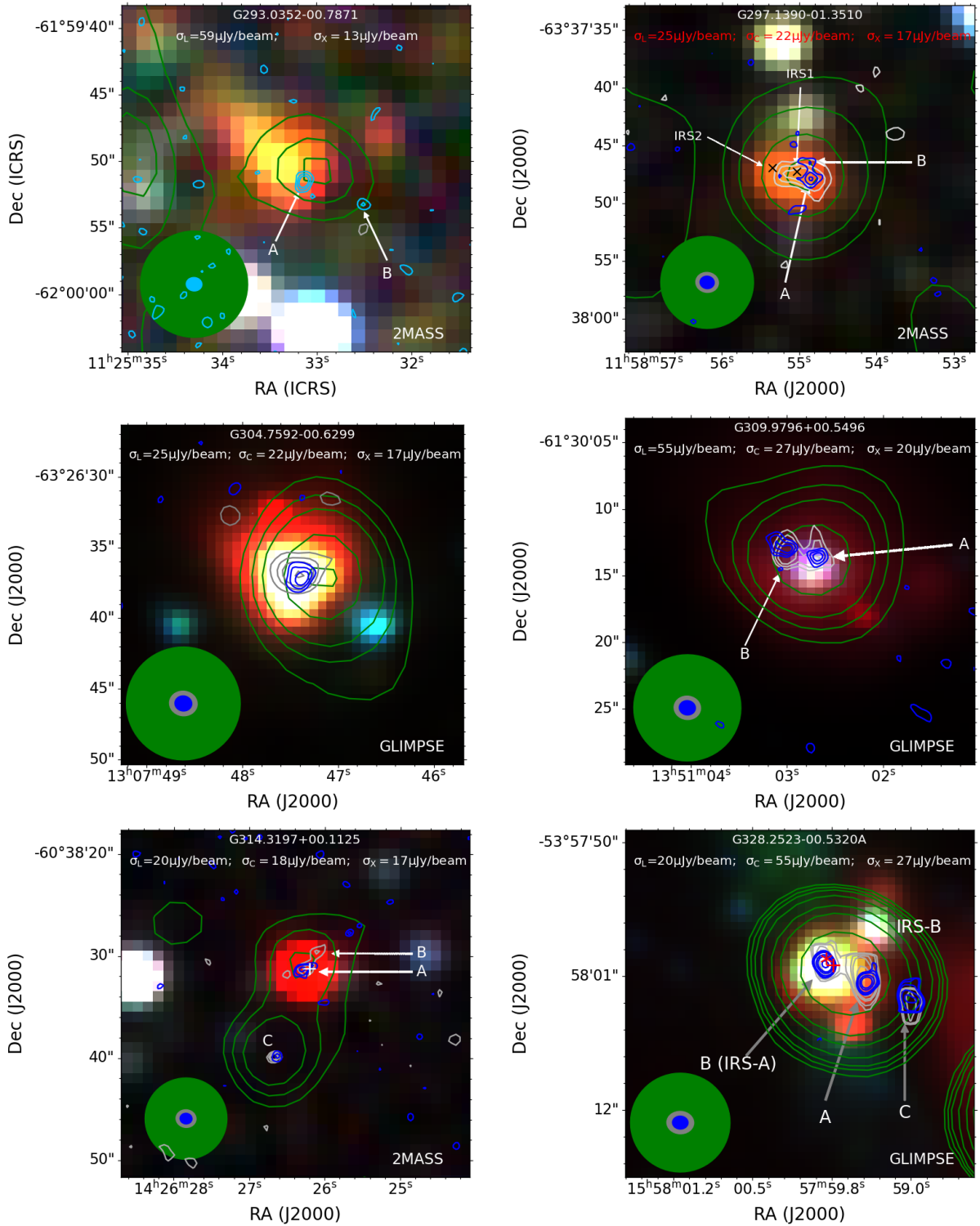

**Figure 1.** continued.

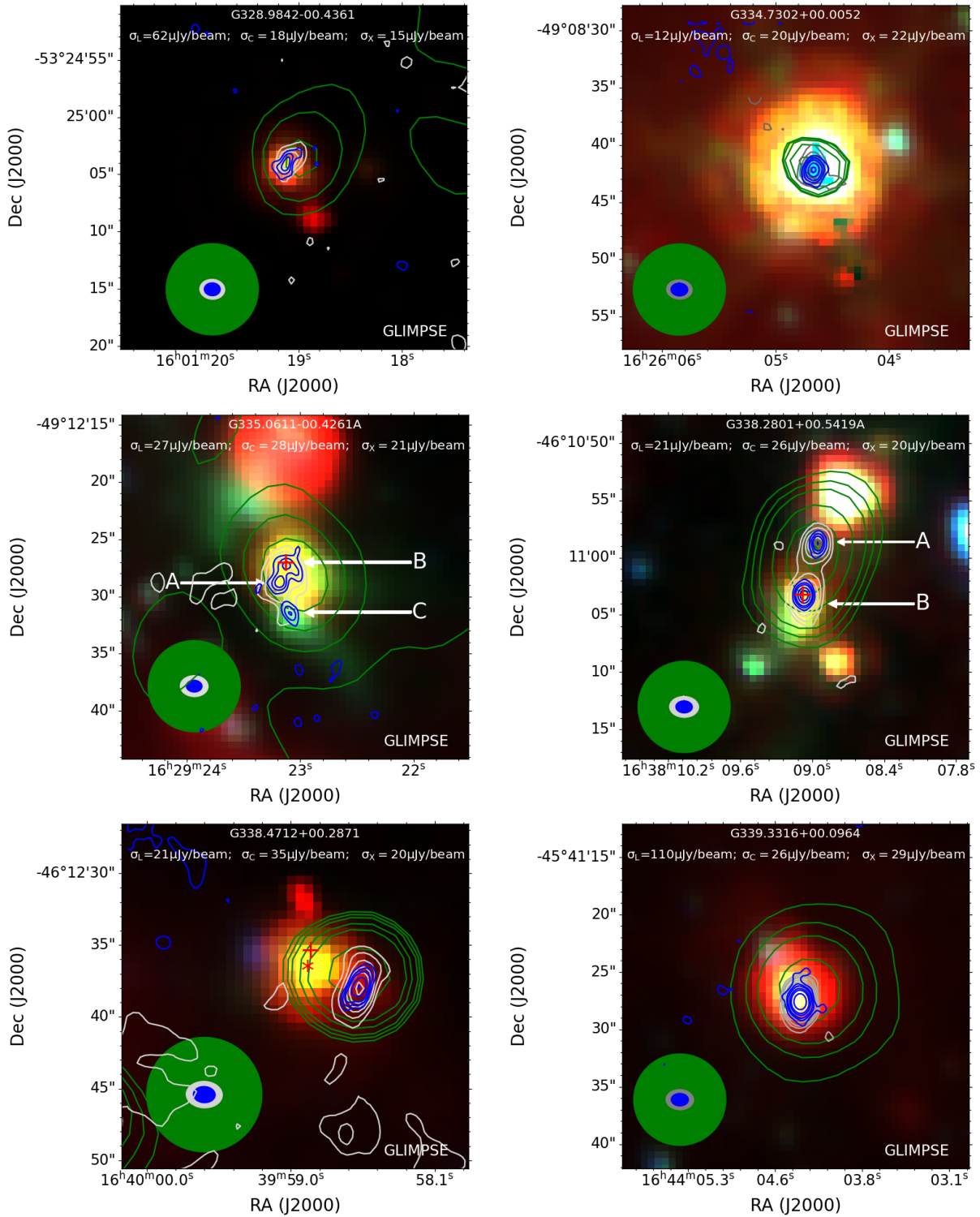

**Figure 1.** continued.

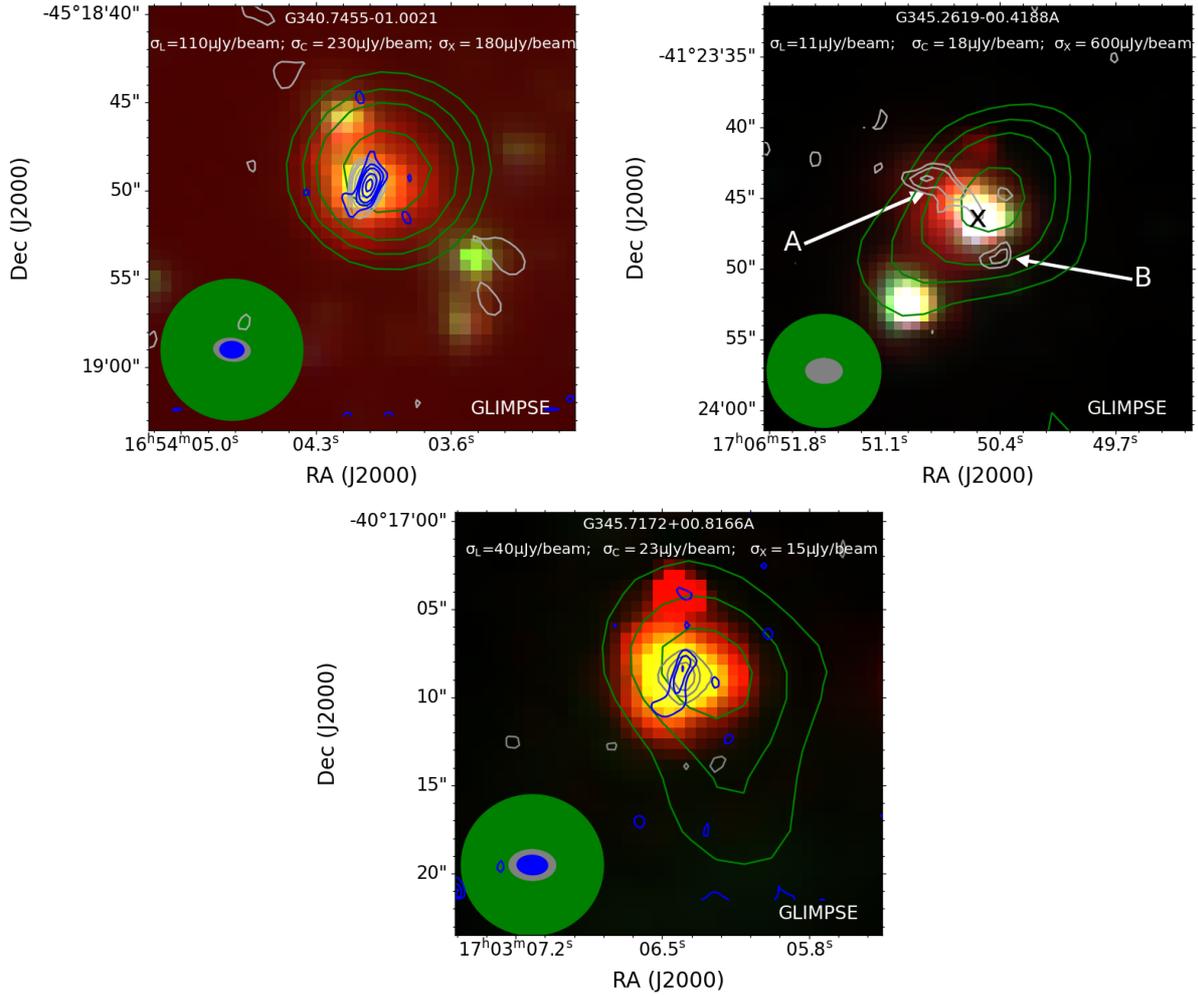

**Figure 1.** continued.

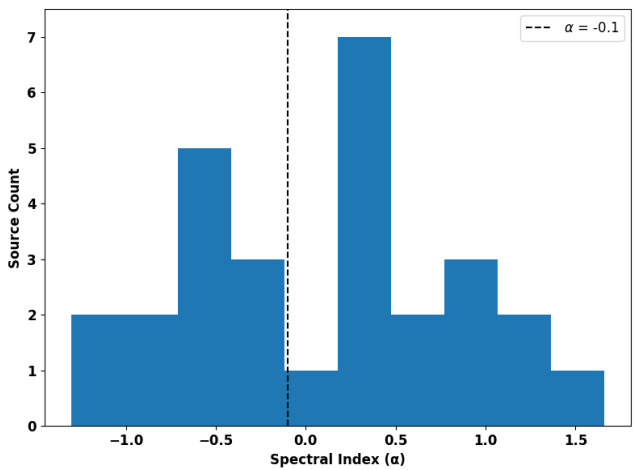

**Figure 2.** A histogram of the spectral indices of the radio sources detected in the ATCA observations. Indices of both the components of the objects, if resolved, as well as the objects are shown. The dashed vertical line at $\alpha = -0.1$ represents the typical dividing line between thermal and non-thermal sources.

### 3.3.3 G282.8969-01.2727

G282.8969-01.2727 was detected at 5.5 GHz but not 9.0 GHz, where the field rms is 0.027 mJy/beam. The emission at 5.5 GHz manifests an extended source consisting of two components, A and B, with flux densities 0.09±0.03 mJy and 0.07±0.02 mJy, respectively.

The spectral index of the extended emission, calculated using the flux at 5.5 GHz, and the upper limit at 9.0 GHz, is <-1.41, indicating non-thermal emission. The steep negative index might result from the resolution of flux at higher frequencies, as evidenced by significant size differences between emissions at 1.3 and 5.5 GHz. The upper limits of the spectral indices of components A and B, estimated using the fluxes at 5.5 GHz and $3\times\sigma$ at 9.0 GHz, are <-0.24 and <0.27 respectively.

Navarete et al. (2015) classified the 2.12 $\mu$m molecular hydrogen emission associated with the source as monopolar, with a PA of 350°, indicating the presence of a shock. Furthermore, the alignment of the morphology of the source at 5.5 GHz with that at 1.3 GHz suggests that it is a jet. The PA of the radio source at 1.3 GHz, 118°±3, is consistent with the PA of the alignment of components A and B, i.e, 124°. However, the difference between the PA of the emission reported by Navarete et al. (2015) and the PA of the radio emission suggests the presence of multiple outflows in the field.

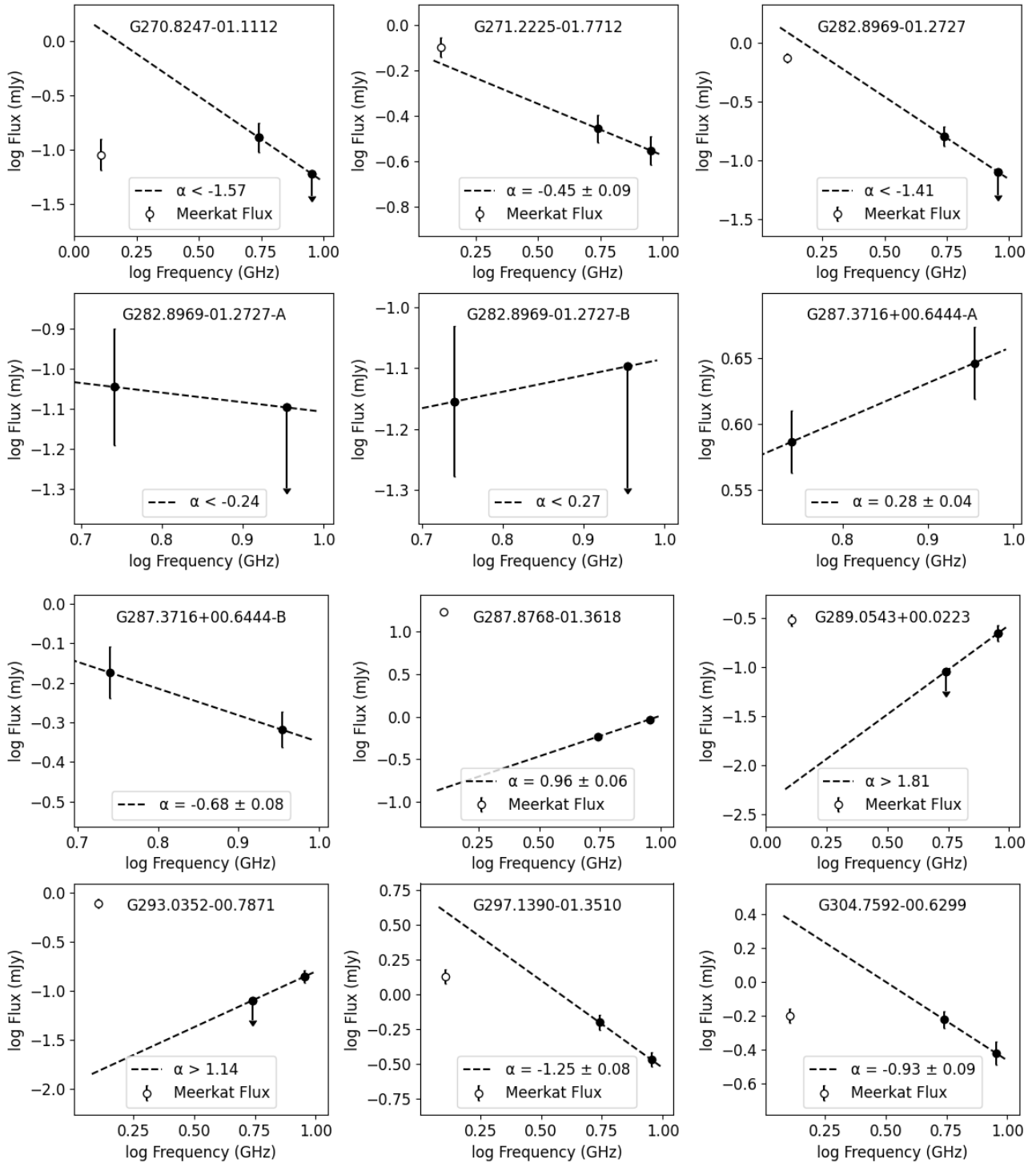

**Figure 3.** Spectral Energy Distributions (SEDs) of detected radio sources (continued on subsequent pages).

While component A exhibits a negative spectral index, its proximity to the methanol maser source (Green et al. 2012) suggests that it could be one of the cores or a lobe that is very close to the core.

#### 3.3.4 G287.3716+00.6444

G287.3716+00.6444 shows two radio sources, A and B, aligned in a NE-SW direction at 5.5 GHz and 9.0 GHz (see Figure 1). The emission observed at 1.3 GHz displays a similar orientation, confirming the alignment of components A and B. The two components were also detected in Purser et al. (2016), further supporting the presence of a jet. The alignment of A and B at a PA ≃ 48° is comparable to the observed PA of the radio source which is 35°±13 (Obonyo et al. 2024).

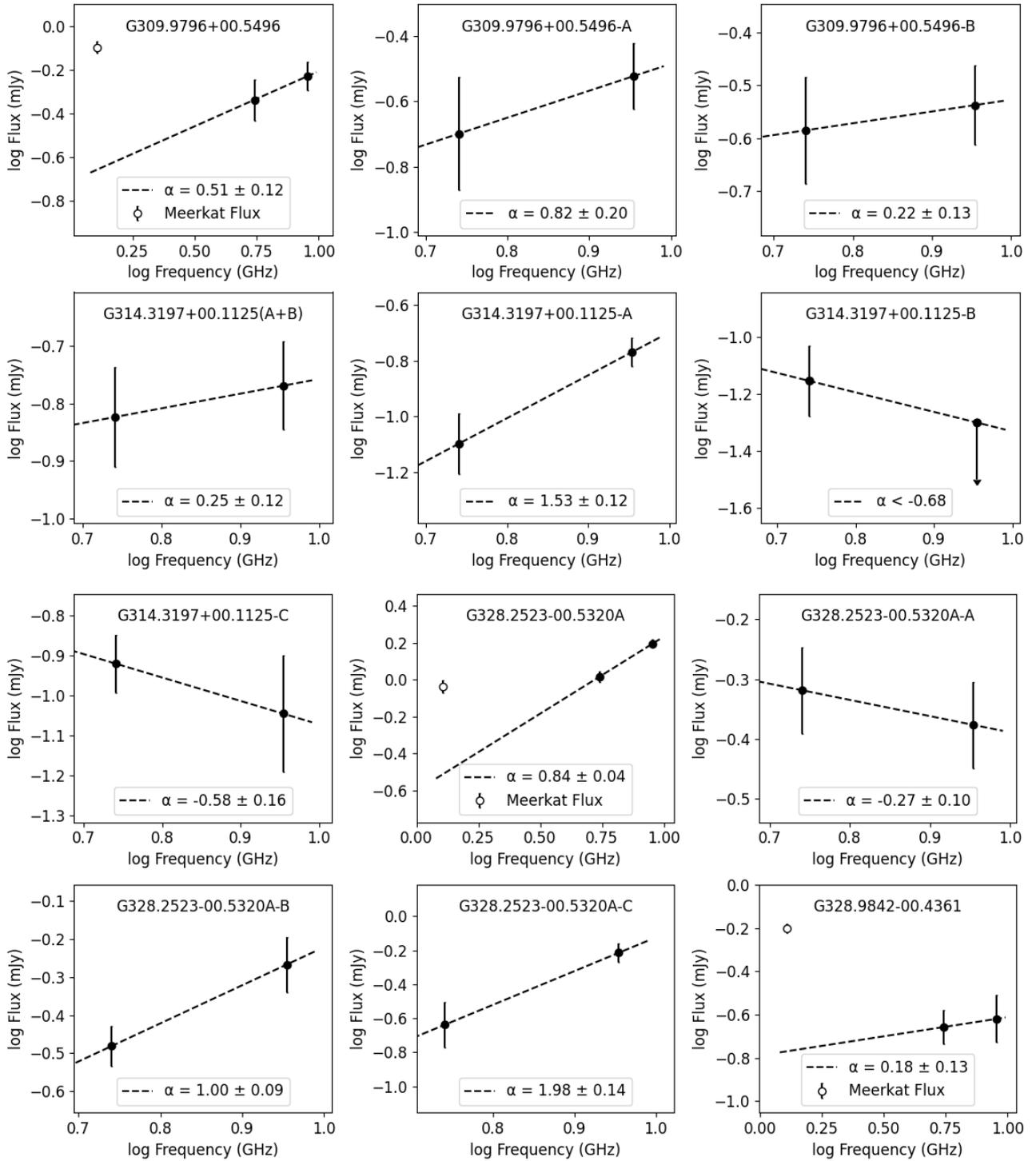

**Figure 3.** Figure 3 (continued).

The spectral index of component A is 0.28±0.04, consistent with ionized winds, while component B exhibits a spectral index of -0.68± 0.08, indicative of non-thermal emission. The indices suggests that A is a thermal core, while B is a non-thermal lobe, implying a core-lobe system. Moreover, the proximity of a methanol maser source identified by Green et al. (2012) to component A reinforces its classification as a core, while B is a likely lobe.

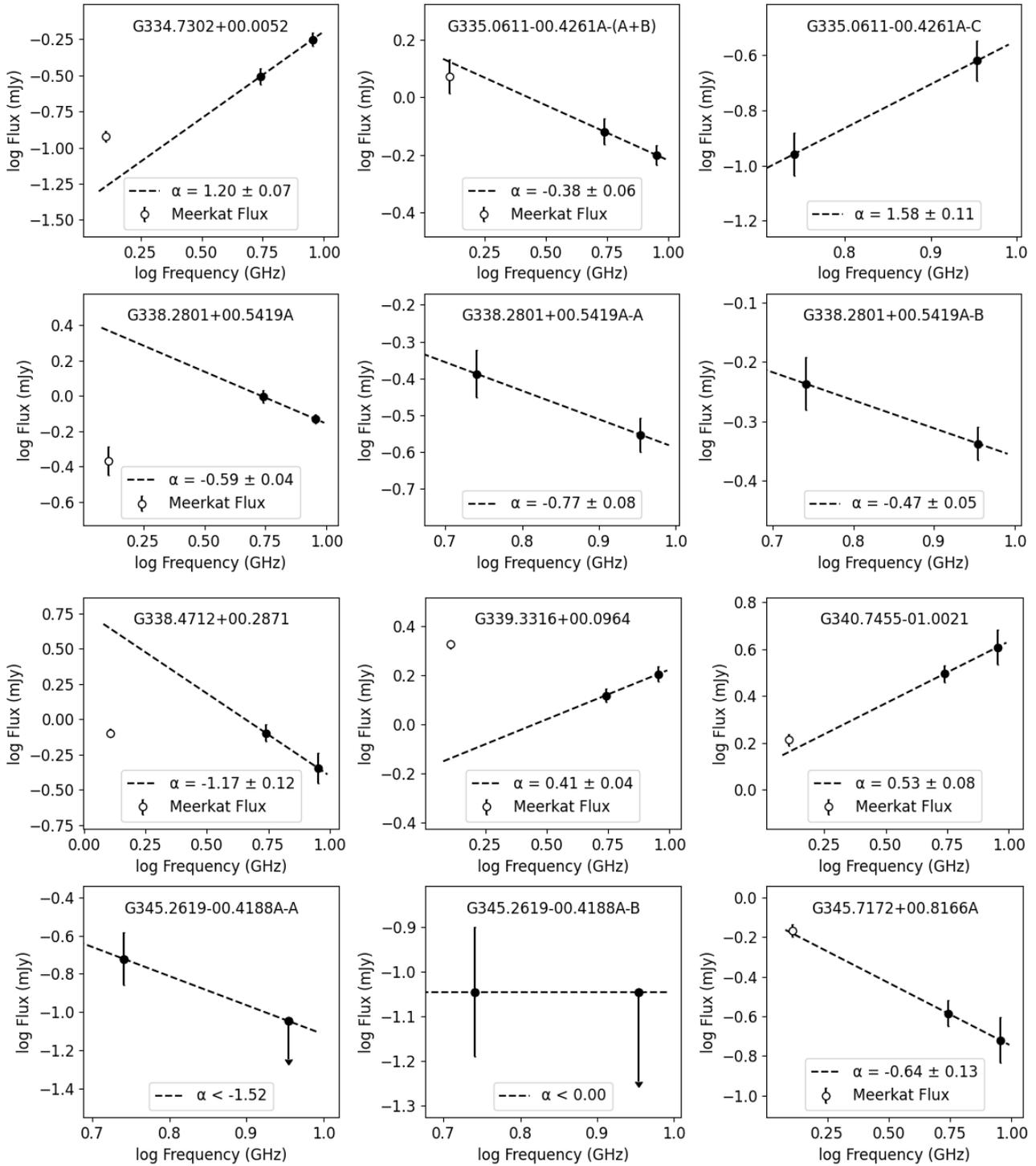

**Figure 3.** Figure 3 (continued).

Both Koumpia et al. (2021) and Ilee et al. (2018) detected Br$_\gamma$ hydrogen recombination line emission at 2.167 $\mu$m, which they interpreted as evidence for the presence of an accretion disk. This suggests that G287.3716+00.6444 is an MYSO that hosts a disk and a jet. However, Purser et al. (2016) classified component A as a compact HII-region, suggesting that it may be a jet-driving HII region, similar to G345.4938+01.4677 (Guzmán et al. 2016).

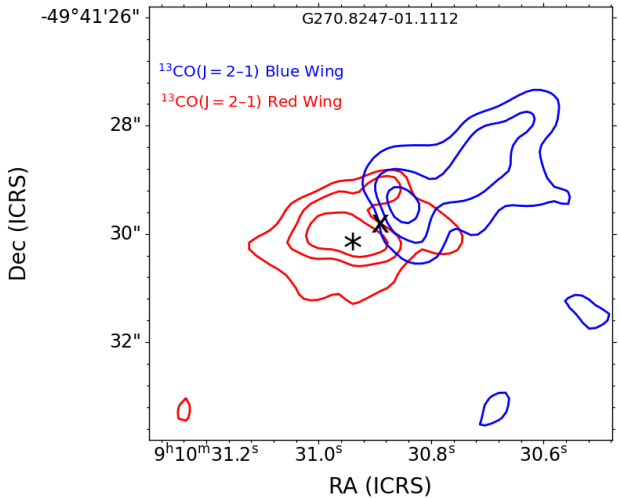

**Figure 4.** Red and blue wings of $^{13}$CO($2 \rightarrow 1$) emission line at 220.39868 GHz in the field of G270.8247-01.1112. The cross (x) and asterisk (*) represent the locations of the IR and mm sources.

### 3.3.5 G287.8768-01.3618

G287.8768-01.3618 exhibits a spectral index of 0.96±0.06, consistent with a thermal radio source. The morphology of its radio emission at 5.5 GHz and 9.0 GHz is aligned in a SE-NW direction, with position angles of 148°±10 and 161°±3, respectively. The consistent orientation of the radio emission and position angles is characteristic of protostellar jets, suggesting that G287.8768-01.3618 is a thermal jet. However, Navarete et al. (2015) detected bipolar emission at 2.122 $\mu$m, a tracer of shock-excited molecular hydrogen at a position angle of 65°. This suggests the presence of another jet propagating along the NE-SW axis, which differs from the orientation of the radio emission. Furthermore, infrared observations by Mottram et al. (2007) detected two point sources aligned with the radio emission in a SE-NW direction at 10.4 $\mu$m. These two infrared sources indicate the potential presence of two independent jet-driving sources in the field. Nevertheless, it remains unclear whether the H$_2$ jet is associated with either of the individual infrared sources as both are aligned with the radio emission.

### 3.3.6 G289.0543+00.0223

G289.0543+00.0223 was detected at 9.0 GHz but not 5.5 GHz, where the rms noise is 0.27 mJy/beam. The spectral index of the source is >1.81, suggesting that it is a thermal radio source. However, its flux at 1.3 GHz is higher than the expected based on the SED derived from 5.5 GHz and 9.0 GHz. This discrepancy suggests either the presence of non-thremal emission or the loss of flux due to resolved-out emission at higher frequencies. The infrared and radio morphologies of the source appear generally spherical, suggesting that it could be a HII region.

### 3.3.7 G293.0352-00.7871

Two radio sources denoted as A and B in Figure 1 were detected within the field of G293.0352-00.7871 at 9.0 GHz but not 5.5 GHz, where the rms noise level is 0.02 mJy/beam. Their flux densities are 0.14±0.02 mJy and 0.08±0.02, respectively. The positions of G293.0352-00.7871 at 1.3 GHz and 9.0 GHz exhibit an offset of ~ 1.4 arcseconds, likely due to the positional accuracy limitations of the SMGPS (Goedhart et al. 2024). Additionally, the radio source shows a spatial offset relative to its associated infrared (IR) counterpart, located to the NE. The 2MASS IR emission from the source displays elongation in the NE-SW direction, comparable to the orientation of a line joining A to B, potentially indicating the presence of an outflow.

The nature of components A and B remains uncertain. Their spectral indices, >1.14 for A and >0.3 for B, suggest that they are thermal sources. However, the flux of source A at 1.3 GHz is higher suggesting that it might be non-thermal. The discrepancy, however, could be attributed to the low resolution of the 1.3 GHz data.

### 3.3.8 G297.1390-01.3510

G297.1390-01.3510 was detected at both 5.5 GHz and 9.0 GHz, exhibiting distinct morphologies at each frequency. At 5.5 GHz, its emission aligns in an E-W direction, whereas at 9.0 GHz, the orientation is in a N-S direction. These orientations are also seen in images with similar uv-range.

The spectral index of the source is -1.25±0.08, indicating that it is a non-thermal source. Additionally, the flux of the source at 1.3 GHz is lower than the expected based on the SED generated using 5.5 GHz and 9.0 GHz observations. This suggests that the index could be closer to -0.6, which is characteristic of synchrotron emission from jets. This non-thermal property of extended emission from massive protostars, as observed in Obonyo et al. (2024), suggests the dominance of the jet emission by unresolved non-thermal lobes within the jet.

Mid-infrared observations by Mottram et al. (2007) detected two sources, IRS1 and IRS2 (see Figure 1). IRS1 is spatially coincident with an MYSO that could be driving the jet responsible for the 5 GHz emission. However, the alignment of radio sources A and B suggests the presence of another outflow oriented in the N-S direction. This indicates the presence of multiple outflows in the field.

### 3.3.9 G304.7592-00.6299

G304.7592-00.6299 has a spectral index of -0.93±0.09, indicating that it is a non-thermal emitter. Its radio emission at 1.28 GHz, 5.5 GHz and 9.0 GHz does not exhibit a clear or consistent orientation of its morphology. However, it manifests a compact source (Mottram et al. 2007) with nebulosity at 8 $\mu$m which is aligned in the NE-SW direction, comparable to the orientation of 1.28 GHz emission. Furthermore, Navarete et al. (2015) did not detect either infrared continuum emission or molecular hydrogen emission at 2.12 $\mu$m. This suggests the absence of an outflow, implying that G304.7592-00.6299 may be an HII region.

### 3.3.10 G309.9796+00.5496

G309.9796+00.5496 exhibits two radio components, A and B, aligned in a NE-SW direction at both 5.5 GHz and 9.0 GHz. Their spectral indices are 0.82±0.20 and 0.22±0.13, respectively, indicating that both components are thermal emitters. The combined spectral index of G309.9796+00.5496 is 0.51±0.12, which is typical of thermal sources. Cyganowski et al. (2008) classified the source as a possible outflow candidate of an Extended Green Object (EGO). The properties of G309.9796+00.5496 suggest that it is a jet, with component A being the potential jet driver due to its proximity to the location of the IR counterpart (Mottram et al. 2007).

### 3.3.11 G314.3197+00.1125

G314.3197+00.1125 is resolved into two radio components at 1.3 GHz, aligned in a SE-NW direction. At 5.5 GHz, the Northwestern component further splits into two components, A and B, maintaining a comparable alignment with PA≃155°. However, only component A was detected at 9.0 GHz. Additionally, a third radio source, denoted as C, was detected south of both A and B at 1.3 GHz, 5.5 GHz and 9.0 GHz, with a BC PA≃166°. The spectral indices of components A, B, and C are 1.53±0.12, <-0.68, and -0.58±0.16, respectively, indicating a bipolar jet morphology with component A as the core and B and C as non-thermal lobes. Moreover, the comparable PAs of AB and BC support the interpretation of a bipolar outflow. The presence of 6.7 GHz methanol maser emission, detected by Green et al. (2012) at the position marked with a plus(+) sign in Figure 1, supports the identification of A as the jet driver.

While the morphology suggests a bipolar jet, the absence of 2.12μm emission in the field (Navarete et al. 2015), the large angular separation of component C from A, and the lack of IR emission associated with C, implies that it could potentially be a non-thermal extragalactic source.

### 3.3.12 G328.2523-00.5320A

G328.2523-00.5320A splits into three components, labelled A, B, and C in Figure 1, which are aligned in a NE-SW direction. These components, detected at both 5.5 GHz and 9.0 GHz, have spectral indices of -0.27±0.10, 1.00±0.09, and 1.98±0.14, indicating that they are non-thermal, thermal, and thermal, respectively. Source B is associated with methanol maser (Green et al. 2012), a dense mm-core (Csengeri et al. 2017), and infrared emission, suggesting that it is the driving source. Components A and C, on the other hand, may be its non-thermal and thermal lobes, respectively.

### 3.3.13 G328.9842-00.4361

G328.9842-00.4361 was detected at both 5.5 GHz and 9.0 GHz, with a spectral index of 0.18±0.13, indicating that it is a thermal jet. However, its SED between 1.3 GHz and 5.5 GHz shows a negative index, $\alpha = -0.73\pm0.04$, suggesting that while the core may be thermal, it might be hosting non-thermal lobes which dominate the emission at lower frequencies. Morphologically, the source exhibits an elongated structure in a SE-NW direction at 1.3 GHz, 5.5 GHz, and 9.0 GHz, a characteristic feature of jets. The position angles of its morphology at these frequencies are 164±8°, 143±21°, and 141±13°, respectively. The consistency in the PAs further supports the interpretation of the source as a jet.

### 3.3.14 G334.7302+00.0052

G334.7302+00.0052 was detected at both 5.5 GHz and 9.0 GHz, with a spectral index $\alpha = 1.20\pm0.07$, consistent with ionized winds, a feature commonly associated with jets driven by MYSOs. While the radio morphology of the source at 1.3 GHz, 5.5 GHz and 9.0 GHz does not exhibit a clear or consistent elongation, the presence of extended 4.5 μm, green emission, similar to the EGOs described by Cyganowski et al. (2008) suggests a potential EGO SW of the source.

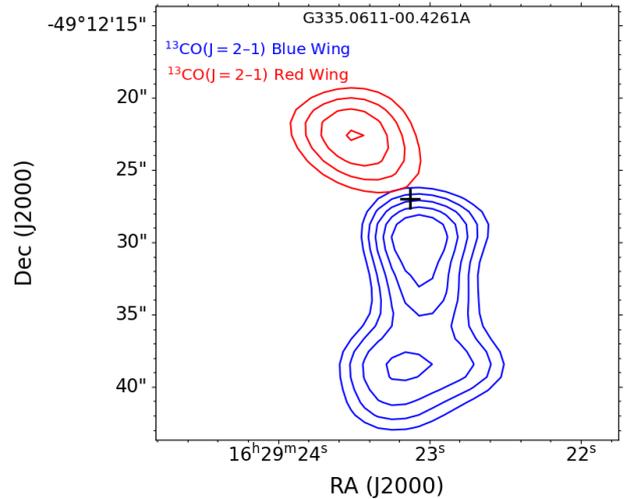

**Figure 5.** Red and blue wings of $^{13}$CO(2 → 1) emission line at 220.39868 GHz in the field of G335.0611−00.4261A. The plus (+) sign represents the locations of methanol maser emission in the field.

The detection of this EGO indicates that G334.7302+00.0052 likely hosts a jet.

### 3.3.15 G335.0611-00.4261A

G335.0611-00.4261A exhibits a complex structure, splitting into two components at 5.5 GHz and three components, A, B, and C, at 9.0 GHz. These components are aligned in a N-S direction, which is consistent with the orientation of the blue wing of $^{13}$CO(2−1) emission observed in the ALMAGAL survey data (see Figure 5). The red wing of the molecular emission is located NE of the associated IR source, aligned along a NE-SW direction.

The orientation of the CO molecular emission are comparable to the morphology of the extended 4.5μm green emission, which traces shock-excited gas (Cyganowski et al. 2008). One of the green emission components extends toward the West, suggesting the presence of multiple outflows in the region. The classification of G335.0611-00.4261A as an EGO by Cyganowski et al. (2008) further supports the interpretation that this source drives a jet.

The spectral indices of the radio components reveal distinct emission mechanisms. Component (A+B) has a spectral index of $\alpha = -0.38\pm0.06$, indicative of non-thermal emission, while component C has a spectral index of $\alpha = 1.58\pm0.11$, suggesting thermal emission. Only B is associated with 6.7 GHz methanol maser emission (Green et al. 2012), suggesting that it may be the core driving the jet, with component A as its lobe. Component C, which has a thermal spectral index, is another potential core.

### 3.3.16 G338.2801+00.5419A

G338.2801+00.5419A is resolved into two radio components, A and B, at both 5.5 GHz and 9.0 GHz in ATCA observations. The morphology observed with both MeerKAT and ATCA aligns in a SE-NW direction, depicting a jet-like core-lobe structure, consistent with the characteristics of a collimated outflow. The spectral index analysis reveals a non-thermal nature for the overall system, with an index of -0.59±0.04. The individual components also exhibit non-thermal characteristics, with component A having an index of -0.77±0.08 and component B -0.47±0.05, indicating that both A and B are non-thermal lobes. Interestingly, component B is associated with 6.7 GHz methanol maser emission (Green et al. 2012), (sub-)mm-core (Elia et al. 2017) and an infrared source, implying that it may either be the driving source of the jet or is located closer to the jet driver. Additionally, a 4.5 $\mu$m green point source is observed to the SE of B, potentially an EGO, further supporting the presence of a bipolar outflow with B as the potential jet driver.

### 3.3.17 G338.4712+00.2871

G338.4712+00.2871 exhibits a jet-like morphology aligned in a SE-NW direction at both 5.5 GHz and 9.0 GHz, with a steep negative spectral index of -1.17±0.12. While this spectral index suggests non-thermal radio emission, its value is atypical. However, the lower 1.3 GHz flux, compared to the SED derived from 5.5 GHz and 9.0 GHz data, implies that the index could be closer to typical value of -0.6.

The morphology of the radio emission is characteristic of jets from massive protostars. Additionally, the position angles of the radio emission at 5.5 GHz and 9.0 GHz are comparable, i.e 171±4 and 157±3, respectively, further supporting the interpretation that the source is a jet. Despite the jet-like characteristics, the radio source does not show association with the 6.7 GHz methanol maser (Green et al. 2012) or infrared sources, which are marked by plus(+) and cross signs in Figure 1, respectively. The maser and infrared sources are located approximately 3.5″ east of the radio source, suggesting that the radio emission is not linked to the infrared source. Thus, the radio emission could be from a highly obscured protostellar object, comparable to a class 0 type protostar, which is undetectable at near-infrared wavelengths (Kern et al. 2016).

The 8 $\mu$m emission associated with G338.4712+00.2871 appears as a point source adjacent to a bright ridge. This source, however, was not detected in 2MASS and TIMMI2 (Mottram et al. 2007) observations, potentially due to strong absorption seen on the 8$\mu$m map, especially towards the west where the radio source is located. This absorption could be obscuring the driving source of the radio jet.

### 3.3.18 G339.3316+00.0964

G339.3316+00.0964 was detected at both 5.5 GHz and 9.0 GHz. The morphologies at these frequencies are elliptical and aligned in a N-S direction, with position angles of 16±6° at 5.5 GHz and 21±46° at 9.0 GHz. The spectral index of the source between 5.5 GHz and 9.0 GHz is 0.41±0.04, indicating that it is a thermal radio source. This value lies between those expected for spherical and collimated ionized winds (Reynolds 1986). However, 1.3 GHz flux suggests a negative spectral index, which is the characteristic of non-thermal emitters, suggesting a combination of thermal and non-thermal emission mechanisms.

The association of the radio source with infrared emission, along with its morphologies at 5.5 GHz and 9.0 GHz and the spectral indices, suggests that G339.3316+00.0964 is a thermal jet. However, the orientation of the infrared emission is misaligned with that of the radio source towards the SW–NE direction. Additionally, the infrared emission exhibits a 4.5 $\mu$m green radiation feature to the NE, suggesting the presence of another jet in that direction or precession.

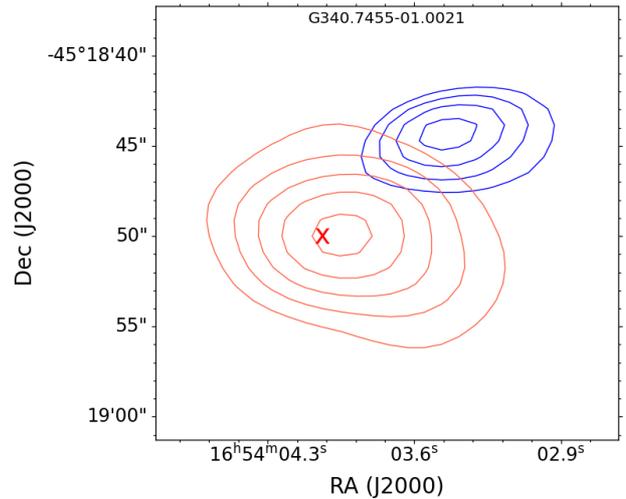

**Figure 6.** Red and blue wings of $^{13}$CO(2 → 1) emission line at 220.39868 GHz in the field of G340.7455−01.0021, shown in red and blue colours, respectively. The plus (X) sign represents the locations of IR source.

### 3.3.19 G340.7455-01.0021

The 5.5 GHz and 9.0 GHz emission from G340.7455-01.0021 displays an extended morphology oriented in a NE-SW direction, similar to alignment of the blue- and red- shifted wings of CO emission in the field (see figure 6). The spectral index of the radio emission is 0.53±0.08, consistent with thermal jets, and the 1.3 GHz emission aligns well with the SED of the jet.

G340.7455-01.0021, classified by Cyganowski et al. (2008) as an EGO outflow candidate, displays 4.5 $\mu$m green emission at a PA≃225°. Additionally, Navarete et al. (2015) detected a single knot at a PA of 290° which may be associated with the radio jet. The difference in position angles of the knots, and radio emission suggests the presence of multiple outflows in the field.

### 3.3.20 G345.2619-00.4188A

G345.2619-00.4188A exhibits two radio components, labelled A and B, detected at 5.5 GHz but not at 9.0 GHz where rms noise is 0.03 mJy. The flux densities of A and B at 5.5 GHz are 0.19±0.06 and 0.09±0.03 mJy respectively. The spectral indices of components A and B are <-1.52 and <0.00, respectively, suggesting that they are non-thermal lobes of a jet.

The alignment of the infrared source with radio components A and B in a NE-SW direction suggests the presence of a bipolar outflow, with component B also exhibiting characteristics of a bow-shock. This jet is likely driven by a core associated with the mid-IR source, whose location is marked with a cross(x) in Figure 1.

### 3.3.21 G345.7172+00.8166A

G345.7172+00.8166A exhibits an extended morphology oriented in a N-S direction, particularly evident at 1.3 GHz and 9.0 GHz, where it displays a characteristic jet-like structure. Its spectral index of -0.64±0.13 is consistent with non-thermal jets. Moreover, the PA of its emission at 5.5 GHz and 9.0 GHz are comparable, i.e, 139±32° and 158±3°, respectively, further supporting the jet-like nature. Cyganowski et al. (2008) classified it as an EGO and an outflow candidate that is associated with an Infrared Dark Cloud (IRDC). This association suggests that G345.7172+00.8166A may be a very young massive protostar.

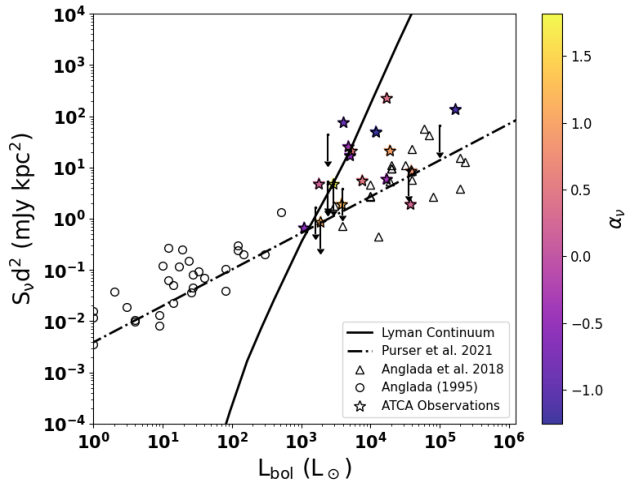

**Figure 7.** Radio continuum luminosity at 5.5 GHz vs bolometric luminosity of the jets in the sample. The asterisks represent the jets in our sample with the face colours showing their spectral indices. The arrows show the upper limits. The solid line represents the radio luminosities expected from Lyman continuum radiation for a zero-age main-sequence star of a given bolometric-luminosity (Thompson 1984). The dashed-dotted line is the least-squares fit to the radio jets reported in Purser et al. (2021).

### 3.4 Radio Luminosity vs Bolometric Luminosity

The radio luminosity of the cores of the objects at 5.5 GHz was calculated and plotted against their bolometric luminosities, taken from Lumsden et al. (2013). For sources with unresolved cores, the fluxes of the jets, which host the cores, were used. This plot is instrumental in distinguishing between free-free emission from collisionally ionizing jets and photoionizing UCHII regions (Anglada et al. 2018). Shock-ionized jets typically exhibit a shallower slope in the radio versus bolometric luminosity plot, whereas photoionized UCHII regions show a steeper dependence due to their strong correlation with the Lyman continuum flux.

Figure 7 shows the plot of radio luminosity versus bolometric luminosity for the objects classified as jets in the sample. Data from previous observations, such as Anglada (1995) and Anglada et al. (2018), are included to illustrate the expected trends. While some sources lie closer to the empirical jet relation from Purser et al. (2021), others are closer to the Lyman continuum line, particularly those with negative spectral indices. Notably, the sources with higher radio luminosities compared to previous studies are predominantly non-thermal. This trend was also observed by Kavak et al. (2021) and Obonyo et al. (2024), who attributed it to differences in the resolution of observations used in estimating the indices. However, the higher radio luminosities exhibited by non-thermal jets compared to thermal counterparts may reflect an intrinsic property of the sources, perhaps an indicator of non-thermal contribution to the free-free emission. Around a bolometric luminosity of $10^3$ $L_\odot$, there is significant overlap between jets and UCH II regions within the observed scatter. This region also corresponds to where zero-age main sequence protostars and massive protostellar jets are expected to lie. In such cases, additional diagnostics such as source morphology or radio recombination line (RRL) widths are essential to distinguish between the two types.

## 4 Conclusions

The higher-resolution ATCA observations of 28 massive young stellar objects, previously unresolved in the SARAO MeerKAT Galactic Plane Survey (Goedhart et al. 2024), at both 5.5 GHz and 9.0 GHz, resulted in 22 significant detections. Among these, 19 were classified as jets, 2 as HII-regions and one as an unknown source. Notably, of the 19 jets, 13 exhibited non-thermal emission or were associated with non-thermal lobes. This indicates that approximately 68% of the jets are associated with non-thermal emission, while the remaining 32% are thermal. These findings suggest that magnetic fields play a crucial role in driving massive protostellar jets (Blandford & Payne 1982, Shu et al. 1994a), supporting the concept of disk-fed accretion similar to that observed in low-mass protostars. However, the mechanism driving thermal jets, especially for creating thermal lobes, remains unclear. The morphology of most of the sources manifests jet-like structures with resolved components aligned linearly, ie. a jet-lobe(s) system. Moreover, molecular outflows also manifested the presence of jets in some cases. The objects were further plotted on a radio- versus bolometric- luminosity plot to eliminate potential UCHII regions. The plot revealed that most of the radio-luminous sources are also non-thermal and are potential jets, implying that their radio luminosity is perhaps due to additional emission from non-thermal radiation.

**Acknowledgements**

Special thanks to the DARA project for sponsoring the observer's trip to CSRIO. Part of the data used here were retrieved from the Japan Virtual Observatory (JVO) portal (http://jvo.nao.ac.jp/portal/) which is operated by ADC/NAOJ.

**Data Availability**

Data Availability Statement: The data underlying this paper are available at the SARAO Data Archive: https://meerkatgps.org/data_products/Mosaic_Moment0/, ATCA archive: https://atoa.atnf.csiro.au and JVO portal: http://jvo.nao.ac.jp/portal/.

This paper has been typeset from a TEX/LATEX file prepared by the author.